\documentclass[amssymb, floatfix, setstack, iopams, amsmath ]{iopart}

\usepackage[margin=1in]{geometry}
\usepackage{graphicx}
\usepackage{dcolumn}
\usepackage{float}
\usepackage{bm}
\usepackage{titlesec}
\usepackage{caption}
\usepackage{subcaption}
\usepackage{fancyhdr}
\usepackage{changepage}
\usepackage{multirow}

\usepackage[colorlinks]{hyperref}

\titleformat{\section}{\bf}{\thesection}{1em}{}

\begin{document}

\noindent \large{\bf Charge-exchange measurements of high-energy fast ions in LHD using negative-ion neutral beam injection} \\

\noindent W.H.J. Hayashi$^1$, W.W. Heidbrink$^1$, C.M. Muscatello$^2$, D.J. Lin$^2$, M. Osakabe$^3$, K. Ogawa$^3$, Y. Kawamoto$^3$, H. Yamaguchi$^3$, R. Seki$^3$, H. Nuga$^3$, M. Isobe$^3$, Y. Fujiwara$^4$, S. Kamio$^{1,4}$

\vspace{0.5\baselineskip}
$^1$Department of Physics and Astronomy, University of California-Irvine, Irvine, CA 92697, USA

$^2$General Atomics, 3550 General Atomics Crt, San Diego, CA 92121, USA

$^3$National Institute for Fusion Sciences, National Institutes of Natural Sciences, Toki 509-5292, Japan

$^4$TAE Technologies, Foothill Ranch, CA 92697, USA

\section*{Abstract}
\begin{adjustwidth}{\parindent}{}
    A new sightline geometry for the fast-ion D-alpha (FIDA) diagnostic on the Large Helical Device (LHD) has been confirmed to measure signals for high-energy fast ions produced by negative-ion neutral beam injection.
    The newly installed sightline uses a 180 keV tangential negative-ion neutral beamline as the active source.
    Due to the small angle between the beamline and FIDA sightline, the relative velocity between fast ions and injected neutrals is small.
    This allows for high-energy fast ions just below the beam injection energy to produce measurable Doppler-shifted FIDA emission.
    Experiments were conducted at LHD in order to compare the new sightline, which views a high-energy negative-ion tangential beamline, and the old sightline, which views a low-energy perpendicular positive-ion  neutral beamline.
    The measured FIDA signal is validated against predictions from the synthetic fast-ion diagnostic code FIDASIM with a distribution function modelled by the 5D transport code GNET.
    The results of the experiment confirm that reducing the viewing angle with a tangential active beam allows FIDA diagnostic to view high-energy fast ions with a sufficient signal-to-noise ratio.
\end{adjustwidth}

\section{\label{sec:one}Introduction}

Many experimental devices in operation today use positive-ion neutral beam injection (PNBI) for heating the plasma but negative-ion neutral beam injection (NNBI) has shown to be operable at higher energies, making it a candidate for use with reactor-relevant scenarios.
NNBI systems have been demonstrated at high energies on the Large Helical Device (LHD) using a 180-keV system \cite{ikeda_first_2018} and JT-60U using a 500-keV system \cite{ohga_present_2002}.
Similar systems are planned for implementation in future devices such as JT-60SA\cite{noauthor_jt-60sa_2018} and ITER\cite{hemsworth_status_2009}.
The performance of NNBI systems is dependent on the heating and current drive provided by fast ions.
Verifying the efficiency of NNBI systems requires an understanding of the fast-ion distribution in the plasma.
Although the fast ion density is often small compared to the thermal plasma density, fast ions account for a significant portion of the stored plasma energy.
Consequently, fast ion transport has a significant effect on confinement as well as overall plasma stability \cite{heidbrink_stability}.

LHD is a stellarator that uses five NBI beamlines for heating hydrogenic plasmas.
Three NNBI beamlines --- NB1, 2, 3 --- are tangentially injected with energies up to 180 keV.
The two additional PNBI beamlines --- NB4, 5 --- are perpendicularly injected with energies up to 80 keV for deuterium and 40 keV for hydrogen.
The NNBI use two negative ion sources per beamline while the PNBI use 4 positive ion sources per beamline for a total of 14 sources.
The different geometries of the beamlines populate different orbits.
The perpendicular beams populate helically trapped orbits while the tangential beams populate passing orbits that travel in orientations analogous to a tokamak's co- and counter-current directions.
Due to the absence of a plasma current in stellarators, the co- and counter- designations instead refer to the direction of $B_t$, the toroidal magnetic field component.
NBI on LHD has used hydrogen since 1998 with a deuterium campaign from 2017 to 2023 \cite{Takeiri_2017}.

\begin{figure}[p]
    \centering
    \includegraphics[width=0.8\columnwidth]{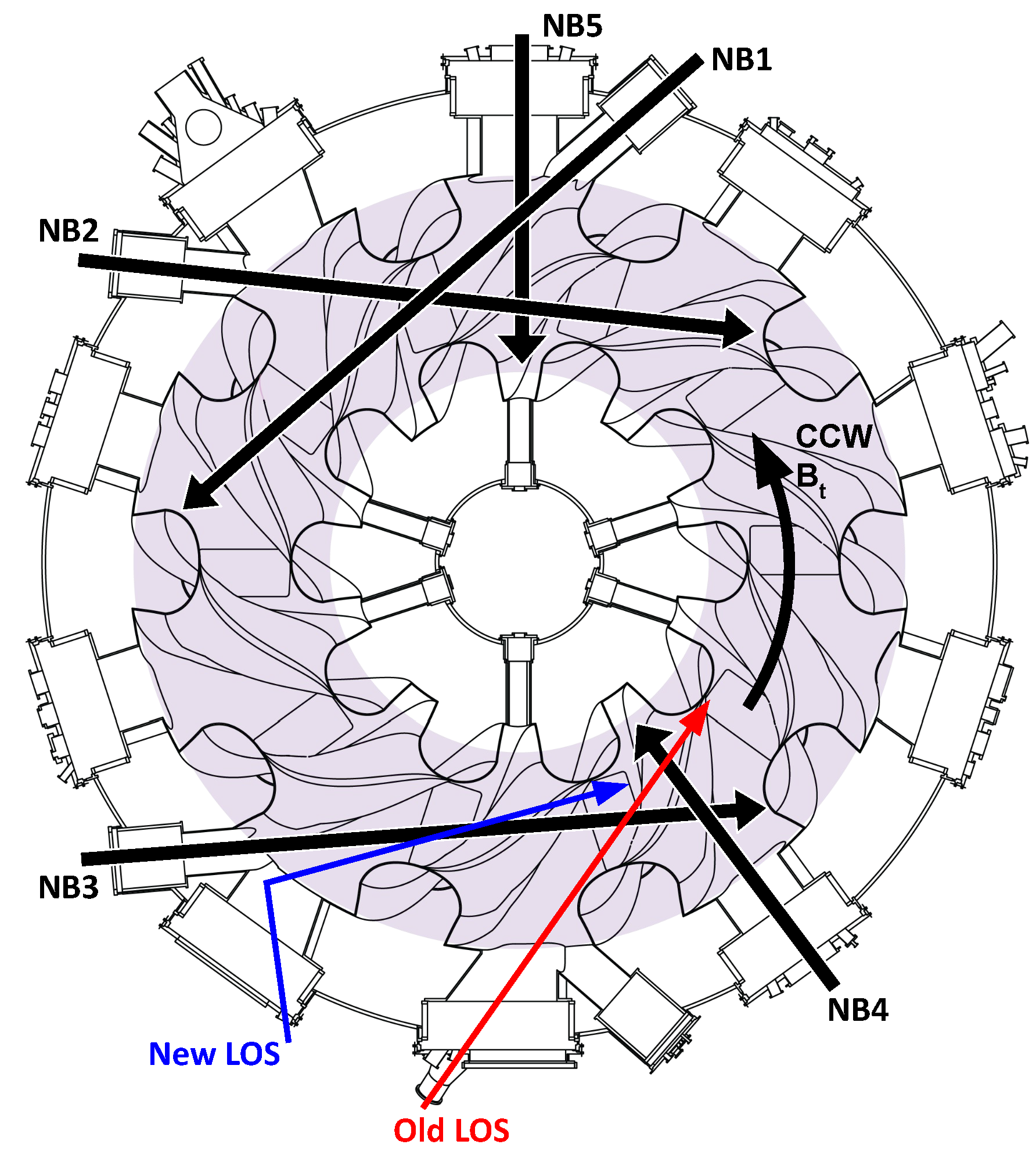}
    \caption{LHD top-down midplane view with beamlines. The high-energy view uses the new LOS (blue) with the negative-ion source NB3 as the active beam while the low-energy view uses the old LOS (red) with the positive-ion source NB4.}
    \label{fig:fig1}
\end{figure}

LHD and other devices in operation today utilize an array of fast-ion diagnostics that are sensitive to different regions of phase-space.
For example, neutron collimators \cite{ogawa_large_2018} are sensitive to high-energy ions but are insensitive to the direction of the velocity including the fast ion’s pitch ($p = v_{||}/v$), where $v_{||} \equiv \vec{v}\cdot\vec{B}/B$ is the component of the fast-ion velocity $\vec{v}$ that is parallel to the magnetic field $\vec{B}$.
Fast-ion D-alpha (FIDA) is a diagnostic that uses charge-exchange recombination spectroscopy to make phase-space resolved measurements of the fast-ion distribution \cite{heidbrink_hydrogenic_2004}.
Typical FIDA systems measure the Doppler-shifted D-alpha emission from energetic neutral deuterium particles that are produced by a charge-exchange reaction between injected neutrals and fast ions in the plasma.
The range of pitch-space that is observed by a particular FIDA sightline is dependent on the geometry with respect to the magnetic field.
The wavelength of the Doppler-shifted signal corresponds to the energy component that is parallel to the diagnostic sightline.
Larger Doppler-shifts are associated with higher energies along the sightline.
Changing the angle between the sightline and the direction of injected neutrals changes the observable wavelength range of the Doppler-shifted FIDA signal.
The direction of the Doppler-shift (red or blue) depends on the direction of the fast ions with respect to the sightline.
For plasmas heated by NB1 or NB3 as shown in Figure \ref{fig:fig1}, the signal is red-shifted; for NB2 the signal is blue-shifted and for NB4/5 both red- and blue-shifted signals are measured.
LHD uses fast-ion charge-exchange spectroscopy (FICXS) \cite{osakabe_fast_2008}, a FIDA variant that diagnoses hydrogenic ions.
The experiment in this paper used deuterium so the FICXS system will be referred to as FIDA.

The charge-exchange cross section for FIDA emission is dependent on the relative energy between the fast ions and the injected neutrals; this dependency peaks at around 30 keV/amu for hydrogenic ions \cite{heidbrink_fast-ion_2010}.
The original FIDA sightline (line-of-sight or LOS) on LHD \cite{fujiwara_fast-ion_2020} is shown in Figure \ref{fig:fig1} with NB4 as the active beam.
A large relative angle between the old LOS with its active beam means that measured signals are limited to relatively low-energy fast ions.
The new LOS uses NB3 as the active beam with a much smaller relative angle.
A small angle means that the relative energy is largely dependent on the relative magnitudes of the injected neutral energy and fast ion energy.
As a result, the FIDA diagnostic with the new LOS is sensitive to fast ions with energies close to the high injected energies from its active beam. 
In this paper, high-energy view will refer to the new LOS with NB3 as the active beam and low-energy view will refer to the old LOS with NB4 as the active beam.

The effectiveness of the high-energy view at diagnosing high-energy fast ions was predicted in a previous study \cite{muscatello_diagnosis_2019} using a forward-modelling approach of an ideal geometry.
The study used the synthetic diagnostic code FIDASIM \cite{geiger_progress_2020} to perform a forward-model analysis of the measured signal.
FIDASIM predicts FIDA and other emissions using inputs of the measured plasma parameters and a fast-ion distribution function provided by a separate code, typically a transport solver code.
The study in this paper uses a distribution provided by the 5-D transport code GNET \cite{Murakami_GNET}.
The signal predicted by FIDASIM is then used to validate the measured FIDA signal.
FIDASIM is also able to calculate weight functions which describe the phase-space sensitivity of a diagnostic sightline.
Using the weight function $W \equiv W(\vec{X}, \vec{V})$, the measured signal is $S$
\begin{equation}
    S = \int_{\vec{X}, \vec{V}} W \ast f d\vec{X} d\vec{V},
\end{equation}
where $f \equiv f(\vec{X}, \vec{V})$ is the fast-ion distribution function and $\vec{X}$ and $\vec{V}$ are position and velocity in phase space \cite{heidbrink_measurements_2007}.

\begin{figure}[p]
    \centering
    \includegraphics[height=0.7\textheight]{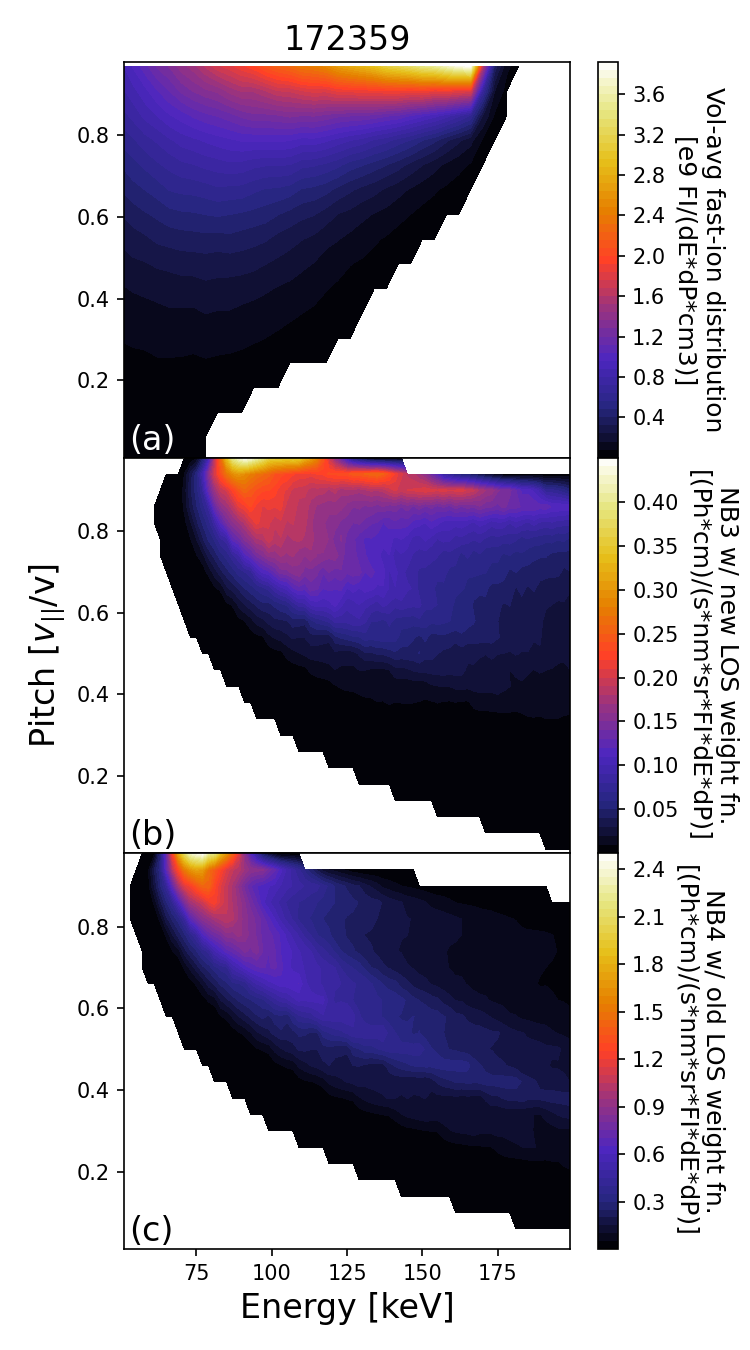}
    \caption{Volume-averaged fast-ion distribution for a typical co-$B_t$ NNBI heated plasma (a). Weight functions for the high-energy view (b) and low-energy view (c) at the intersection with their active beams and the wavelengths of the expected FIDA peak for each view.}
    \label{fig:fig2}
\end{figure}

Weight functions with different phase-space sensitivities facilitate inference of the fast-ion distribution through tomographic inversion \cite{salewski_thesis}.
In particular, the high-energy view is far less sensitive to low-energy fast ions than the low-energy view.
Figure \ref{fig:fig2} shows the fast-ion distribution and weight functions for a typical LHD plasma heated by co-$B_t$ NNBI with a volume-averaged electron density $\tilde{n}_{e} \sim 0.8 \times 10^{19}$ $m^{-3}$ and electron temperature $T_{e0} \sim 5$ keV.
The spatial location of the weight functions correspond to the intersection between each LOS and their respective active beam.
While the position in velocity space corresponds to the predicted wavelength of the FIDA peak.
A comparison of the two weight functions shows that the high-energy view has a sensitivity to higher energies with a peak above 85 keV while the low-energy view is sensitive to lower energies with a peak around 75 keV.

The paper is structured as follows: the setup for the experiment and method for selecting cases is detailed in Sec. \ref{sec:two}, the analysis and results are discussed in Sec. \ref{sec:three}, and Sec. \ref{sec:four} concludes the paper.

\section{\label{sec:two}Experimental Conditions}

\begin{figure}[p]
    \centering
    \includegraphics[height=0.45\textheight]{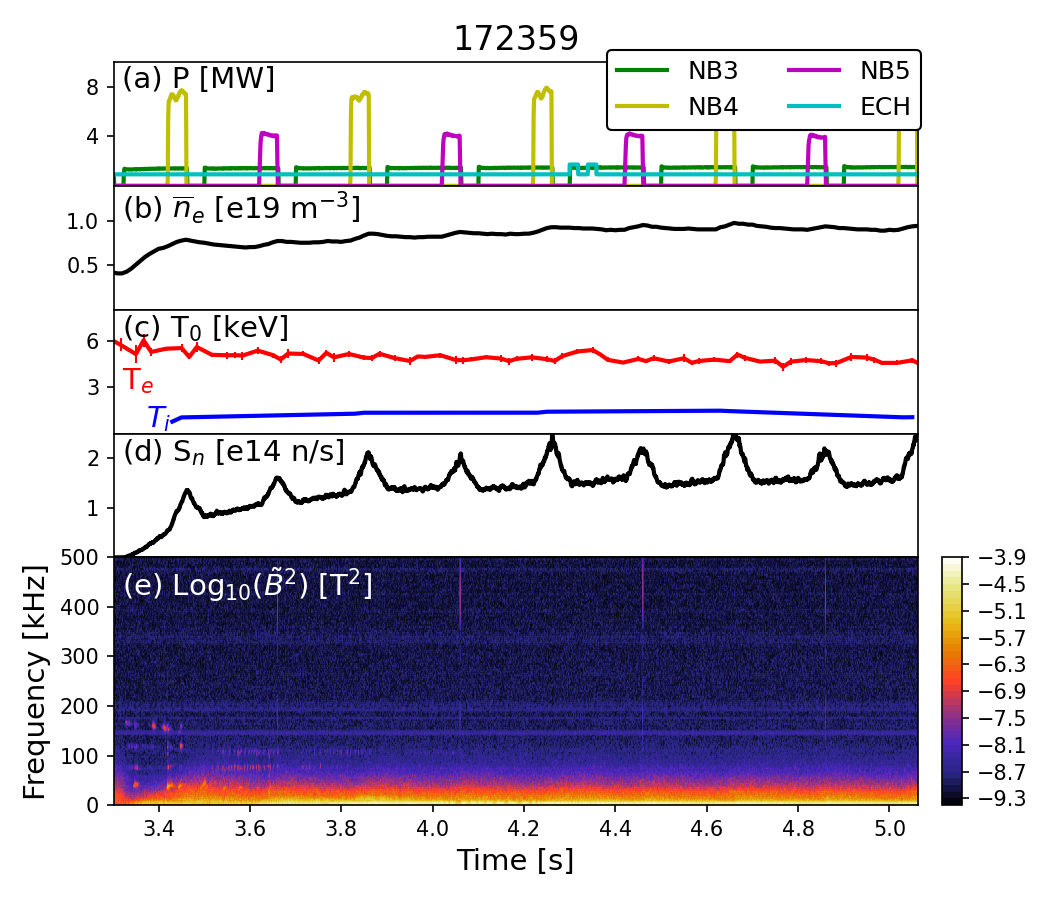}
    \caption{Time series for an MHD-quiescent case showing auxiliary heating power (a), volume-averaged electron density (b), central electron and ion temperatures (c), neutron emission rate (d), and  magnetic fluctuation spectrogram (e).}
    \label{fig:fig3}
\end{figure}

Low-density deuterium experiments were conducted on LHD with the high-energy view; the low-energy view was also used for comparison.
Background-subtracted FIDA signals were measured using modulated beam patterns.
NNBI: modulated with an 80\% duty cycle.
PNBI: modulated with a 20\% duty cycle.
The large duty cycle beam pattern for NNBI allows the beam to populate the fast ions, while also serving as an active diagnostic beam for charge-exchange measurements.
The injection energies and powers of NNBI and PNBI are 166 keV with 1.5 MW and 56 keV with 3.5 MW, respectively.
The experiment used a magnetic configuration with a counterclockwise field B = 2.75 T and magnetic axis at $R_{ax} = 3.6$ m.
The plasma had a central electron temperature of $T_{e0} = 5$ keV, central ion temperature of $T_{i0} = 1.4$ keV, a volume-averaged electron density of $\tilde{n}_{e} = 0.8 \times 10^{19}$ $m^{-3}$, and a counterclockwise plasma current of $I = 5$ kA.
The beam pattern and plasma parameters for a typical case are shown in Figure \ref{fig:fig3}.
Instabilities are observed throughout the experiment.
Additionally, the low-frequency noise ($<$ 15 kHz) seen in Figure \ref{fig:fig3}(e) is observed in every case.
A few quiescent plasmas are selected for comparison with FIDASIM.

\begin{figure}[p]
    \centering
    \includegraphics[height=0.4\textheight]{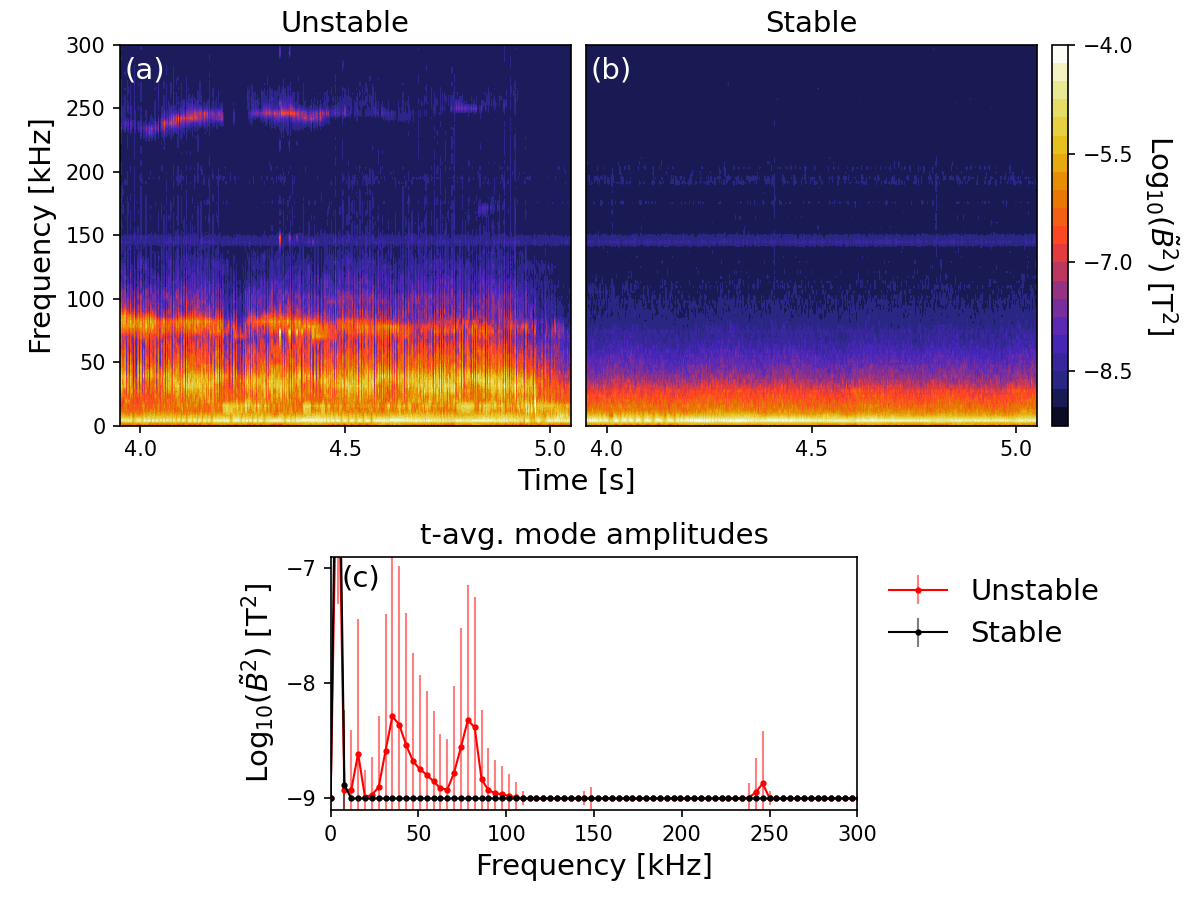}
    \caption{Magnetic fluctuation spectrograms for an unstable case (a) and a stable case (b). Time-averaged instability mode amplitudes (c) for both cases with an artificial noise floor set to $1\times10^{-9}$.}
    \label{fig:fig4}
\end{figure}

The simulated fast-ion distribution function that is input to FIDASIM is calculated for steady state with no instabilities.
MHD-quiescent cases are selected using quantitative analysis of the measured magnetic field fluctuations.
From this study, 81 out of 213 cases are marked as stable.
Figure \ref{fig:fig4} shows an example comparison of an unstable (a) and stable (b) case as well as the time-averaged mode amplitudes (c).
The mode amplitudes show that the stable case only has low frequency activity ($<$ 40 kHz) while the unstable case shows multiple large-amplitude peaks below 100 kHz as well as a peak around 250 kHz.

The experiment compared the high- and low-energy views using the LHD FIDA system to acquire simultaneous measurements.
The FIDA system on LHD consists of 400 $\mu$m optical fibers and a FLP-200 BUNKOKEIKI spectrometer with a grating number of 1200 mm, focal length of 200 mm, and F-number of 2.8.
The signal is measured on a iXon 897 ANDOR electron multiplying charge-coupled device.
16 channels were divided between the new and old LOS with 10 fibers directed to the new LOS and 6 fibers to the old LOS.
Table \ref{table:table1} shows the radial location of the intersections and the relative angles of each channel with the relevant active beam.

\section{\label{sec:three}Experimental Results}

\begin{figure}[p]
    \centering
    \includegraphics[width=0.95\textwidth]{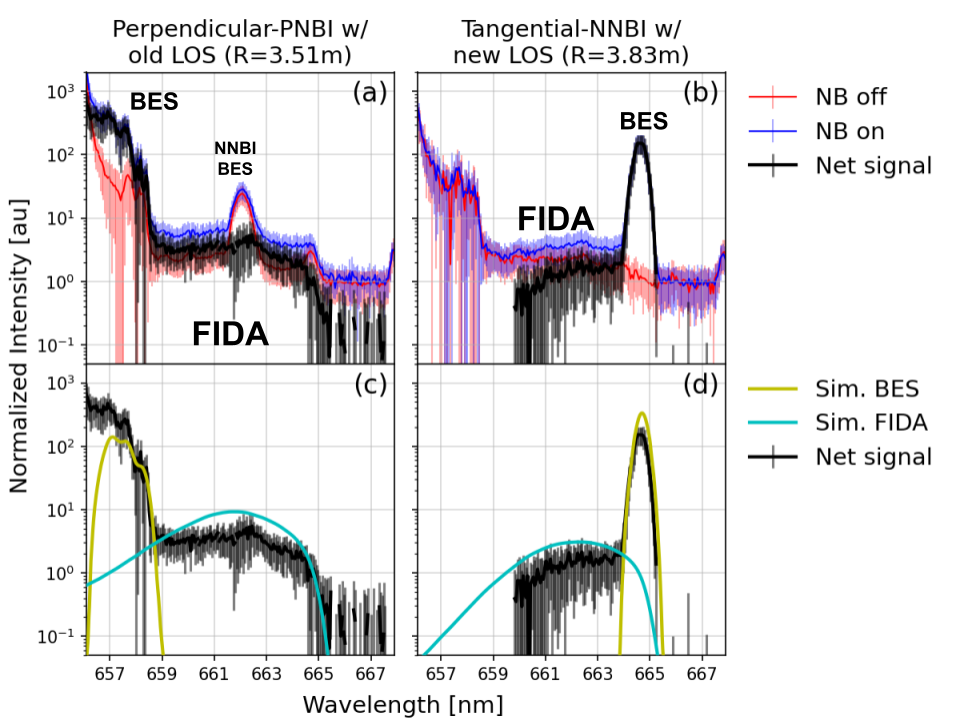}
    \caption{Background-subtracted net signal with active NB on and off (a, b) and simulated FIDA and BES (c, d). Error bars represent 1 standard deviation. NNBI BES can be seen in (a) at a lower Doppler-shift compared to the BES signal in (b).}
    \label{fig:fig5}
\end{figure}

The FIDA emission observed in the selected MHD-quiescent case 172359 shows signals that are Doppler-shifted to the expected wavelength for high-energy fast ions.
Background subtraction is used to construct the net signal using the difference between the beam-on time t = 3.98 s and the beam-off time t = 3.88 s.
Figure \ref{fig:fig5} shows the time-averaged beam-on, beam-off, and net signal normalized to the background bremsstrahlung for the low-energy (a) and high-energy (b) views.
The signals were initially normalized to the beam emission spectroscopy (BES) feature.
However, the high intensity BES signals were found to be truncated above a certain count number making it unsuitable for normalization.
The net signal in \ref{fig:fig5}(b) shows the NNBI BES feature around 665 nm, as expected for a 166 keV injection energy, and the FIDA emission feature between 660 nm and 664 nm.
The peak in the FIDA emission occurs at 662.2 nm which corresponds to fast ions with an energy component of 81 keV along the sightline.
A similar analysis of the low-energy view found that the FIDA peak appears at 661.8nm or 71 keV along the sightline.
The high-energy view has a greater Doppler-shift of the FIDA peak and a reduced signal for lower energies indicating that the high-energy view can isolate fast ions with greater energies.

The behavior of the net signal measured by the high-energy view is validated against FIDASIM under the same conditions as the experimental case.
The comparison of the synthetic signals are shown in Figure \ref{fig:fig5}(c) for the low-energy view and \ref{fig:fig5}(d) for the high-energy view.
Instrumental broadening was applied to the FIDASIM signal using empirically determined values from matching the OV impurity lines.
The shape of the synthetic FIDA features have good agreement with the measured signal between 660 nm and 663 nm, while the normalized magnitude for both views is overestimated by FIDASIM.
This discrepancy can be attributed to the loss of fast ions in the experiment due to charge-exchange with cold neutrals, among other considerations which are not accounted for in the simulation.

\begin{figure}[p]
    \centering
    \includegraphics[height=0.6\textheight]{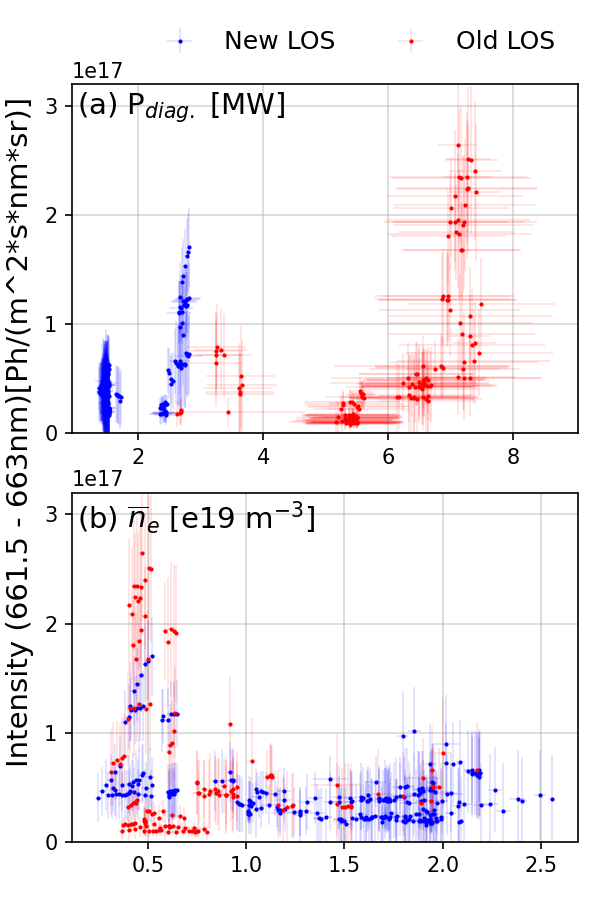}
    \caption{Wavelength-averaged net FIDA signal (661.5 - 663nm) vs injected power of the relevant diagnostic beam (a) and volume-average electron density (b). Error bars represent 1 standard deviation.}
    \label{fig:fig6}
\end{figure}

The parametric dependencies of the FIDA feature are in qualitative agreement with theory.
Theoretically, in addition to the cross-section dependence, the FIDA signal is proportional to the product of the injected-neutral and fast-ion densities, $n_{inj}n_f$.
Increasing power of the diagnostic beam tends to increase $n_{inj}$.
For the beam patterns utilized in this experiment, the diagnostic beams also contribute to the measured fast-ion density.
Figure \ref{fig:fig6}(a) shows that, as expected, the signals for both the low- and high-energy views tend to increase with diagnostic beam power.
Owing to the higher injected power of the PNBI compared to NNBI, the low-energy view has signals that are about twice as large as the high-energy view.
Figure \ref{fig:fig6}(b) shows that both the old and the new signals tend to decrease with increasing density $\bar n_e$.
This also is expected \cite{fujiwara_fast-ion_2020,Luo_FIDA}, since increasing density tends to decrease beam penetration (reducing $n_{inj}$) and increasing density lower the slowing-down time (reducing $n_f$).

\section{\label{sec:four}Conclusion}

Experimental results confirm the predictions that FIDA sightlines nearly tangential to the active beamline can be used to measure high-energy fast ions.
The study in \cite{muscatello_diagnosis_2019} used an ideal geometry for the high-energy view with a 180 keV NNBI heated plasma.
This results in a predicted FIDA peak corresponding to 662.5 nm or fast ions with energies of 90 keV along the sightline.
The difference in energy between the prediction in \cite{muscatello_diagnosis_2019} and the experimental results can be attributed to using a lower injection energy (166 keV) and a deviation of the installed sightline from the ideal geometry.
Despite this, the high-energy view on LHD can still isolate signals from fast ions with over 10 keV higher energies than the low-energy view.
Based on this method, FIDA emission can be measured for fast ions above 100 keV in NBI-heated plasmas using higher injection energies.

\section*{\label{sec:data}Data availability}

The data from this study is available on the LHD experiment data repository \cite{noauthor_lhd_nodate}.

\section*{\label{sec:ack}Acknowledgements}

This study was supported by DOE DE-SC0022131 and DE-SC0018255.
The experiments and analysis were conducted with the support of the LHD team.

\section*{References}
\bibliographystyle{vancouver}
\bibliography{references}

\pagebreak

\clearpage

\begin{table}[p]
\begin{center}
    \begin{tabular}{c c c c}
    \hline
    & Ch & R [m] & $\theta$ [$^{\circ}$] \\
    \hline
    \multirow{10}{3em}{High-energy view} & 1  & 3.94 & 10 \\
    & 2  & 3.94 & 10 \\
    & 3  & 4.12 & 9  \\
    & 4  & 4.06 & 9  \\
    & 5  & 4.00 & 10 \\
    & 6  & 3.96 & 10 \\
    & 7  & 3.94 & 10 \\
    & 8  & 3.93 & 10 \\
    & 9  & 3.83 & 11 \\
    & 10 & 3.71 & 12 \\
    \\
    \multirow{6}{3em}{Low-energy view} & 11 & 3.24 & 33 \\
    & 12 & 3.33 & 34 \\
    & 13 & 3.42 & 35 \\
    & 14 & 3.51 & 36 \\
    & 15 & 3.60 & 36 \\
    & 16 & 3.69 & 37 \\
    \hline
    \end{tabular}
    \caption{Channel number for each diagnostic sightline, the radial location of the LOS-beam intersection, and the LOS-beam angle.}
    \label{table:table1}
\end{center}
\end{table}

\end{document}